\documentclass{ws-procs10x7}

\newcommand{\Bbar}{\,\overline{\!B}}
\newcommand{\Dbar}{\,\overline{\!D^0}}

\newcommand{\BsorBsbar}{\raisebox{7.7pt}{$\scriptscriptstyle(\hspace*{8.5pt})$}
  \hspace*{-10.7pt}\!\Bbar_{s}}

\newcommand{\DorDbar}{\raisebox{7.7pt}{$\scriptscriptstyle(\hspace*{11.5pt})$}
  \hspace*{-13.7pt}\!\Dbar}

\newcommand{\lt}{\left}
\newcommand{\rt}{\right}
\newcommand{\nn}{\nonumber \\}
\newcommand{\no}{\nonumber}
\newcommand{\ov}{\overline}
\newcommand{\eq}[1]{Eq.~(\ref{#1})}
\newcommand{\lqcd}{\Lambda_{\textit{\scriptsize{QCD}}}}
\newcommand{\lhad}{\Lambda_{\textit{\scriptsize{had}}}}
\newcommand{\epm}[2]{
 \raisebox{-0.5ex}{\shortstack[l]{$\scriptstyle+#1$\\$\scriptstyle-#2$}}
                    }
\newcommand{\bra}[1]{\langle \, #1 \, | }
\newcommand{\ket}[1]{| \, #1 \, \rangle }
\newcommand{\imag}{{\rm Im}\,}

\newcommand{\ds}{\displaystyle}
\newcommand{\tev}{\,\mbox{TeV}}
\newcommand{\gev}{\,\mbox{GeV}}
\newcommand{\mev}{\,\mbox{MeV}}

\newcommand{\bbms}{$\mathrm{B_s}\!-\!\ov{\mathrm{B}}{}_\mathrm{s}\,$\ mixing}
\newcommand{\bbmd}{$\mathrm{B_d}\!-\!\ov{\mathrm{B}}{}_\mathrm{d}\,$\ mixing}

\newcommand{\bbm}{$\mathrm{B}\!-\!\ov{\mathrm{B}}{}\,$\ mixing}
\newcommand{\kkm}{$\mathrm{K}\!-\!\ov{\mathrm{K}}{}\,$\ mixing}

\newcommand{\dm}{\ensuremath{\Delta m}}

\newcommand{\fig}[1]{Fig.~\ref{#1}}

\newcommand{\prd}{{\it Phys.\ Rev.~} D}

\newcommand{\rd}{}
\newcommand{\gn}{}

\begin{document}

\title{TTP05-25\hfill hep-ph/0511125\\[5mm]
\large Quark mixing and CP violation --- the CKM matrix}

\author{Ulrich Nierste}

\address{
        Institut f\"ur Theoretische Teilchenphysik,
        Universit\"at Karlsruhe,
        76128 Karlsruhe,
        Germany\\
E-mail: nierste@particle.uni-karlsruhe.de
}

\twocolumn[\maketitle\abstract{I present the status of the elements and
  parameters of the Cabibbo-Kobayashi-Maskawa (CKM) matrix and summarise
  the related theoretical progress since \emph{Lepton-Photon 2003}. 
  One finds 
  $|V_{us}| = 0.2227\pm 0.0017$ from K and $\tau$ decays and 
  $|V_{cb}| = (41.6 \pm 0.5) \cdot 10^{-3}$ from inclusive semileptonic 
  $B$ decays. The unitarity triangle can now
  be determined from tree-level quantities alone and the result agrees
  well with the global fit including flavour-changing neutral current
  (FCNC) processes, which are sensitive to new physics. From the global
  fit one finds the three CKM
  angles $\theta_{12}=12.9^\circ\pm 0.1^\circ$,  
  $\theta_{23}= 2.38^\circ\pm 0.03^\circ$ and 
  $\theta_{13}= 0.223^\circ\pm 0.007^\circ$ 
  in the standard PDG convention. The CP phase
  equals $\delta_{13}\simeq \gamma =  
   (58.8^{\,+ 5.3}_{\,- 5.8})^\circ $ at 1$\sigma$ CL and 
   $\gamma=(58.8^{\,+ 11.2}_{\,- 15.4})^\circ $ at 2$\sigma$ CL.
   A major progress are first results from fully unquenched lattice QCD
   computations for the hadronic quantities entering the UT fit. 
   I further present the calculation of three-loop QCD
   corrections to the charm contribution in $K^+\to \pi^+ \nu \ov{\nu}$
   decays, which removes the last relevant theoretical uncertainty from
   the $K\to \pi \nu \ov{\nu}$ system. Finally I discuss mixing-induced
   CP asymmetries in $\rd b\to s \ov{q} q$ penguin decays, whose naive
   average is below its Standard Model value by 3$\sigma$.  }  ]

\section{Flavour in the Standard Model}%
In the Standard Model transitions between quarks of different
generations originate from the Yukawa couplings of the Higgs field to
quarks. The non-zero vacuum expectation value $v$ of the Higgs field
leads to quark mass matrices $M^u$ and $M^d$ for the up-type and
down-type quarks, respectively.  The transformation to the physical mass
eigenstate basis, in which the mass matrices are diagonal, involves
unitary rotations in flavour space.  The rotation of the left-handed
down-type quarks relative to the left-handed up-type quarks is the
physical Cabibbo-Kobayashi-Maskawa (CKM) matrix $V$. It appears in the
couplings of the $W$ boson to quarks and is the only source of
transitions between quarks of different generations.  $V$ contains one
physical complex phase, which is the only source of CP violation in
flavour-changing transitions.

Flavour physics first aims at the precise determination of CKM elements
and quark masses, which are fundamental parameters of the Standard
Model. The second target is the search for new physics, pursued by 
confronting high precision data with the predictions of the Standard
Model and its extensions.  To this end it is useful to distinguish
between charged-current weak decays and flavour-changing neutral current
(FCNC) processes.  The determination of CKM elements from the tree-level
charged-current weak decays, discussed in Sect.~\ref{sect:tree}, is
practically unaffected by possible new physics.\footnote{ Still new
  physics can be revealed if the $3\times 3$ CKM matrix $V$ is found to
  violate unitarity: One may then infer the existence of new (for example
  iso-vector) quarks which mix with the known six quarks. Further
  leptonic decays of charged mesons are tree-level, but sensitive to
  effects from charged Higgs bosons.}  By contrast, FCNC processes are
very sensitive to virtual effects from new particles with masses at and
above the electroweak scale, even beyond 100 \tev\ in certain models of
new physics. FCNC processes are discussed in Sect.~\ref{sect:loop}.

\begin{figure}
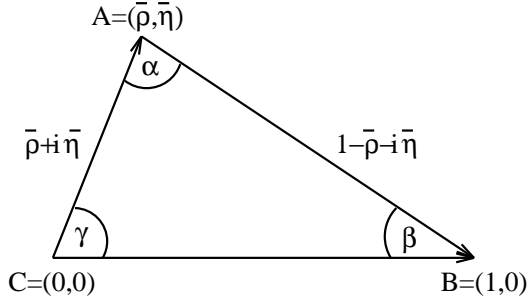

\epsfxsize\columnwidth
\figurebox{}{}{triangle.ps}
\caption{Unitarity triangle (UT).}
\label{fig:ut}
\end{figure}

$V$ can be parameterised in terms of three mixing angles $\theta_{12}$,
$\theta_{23}$, $\theta_{13}$ and one complex phase $\delta_{13}$, which
violates CP . Adopting the PDG convention\cite{pdg}, in which $V_{ud}$,
$V_{us}$, $V_{cb}$ and $V_{tb}$ are real and positive, these parameters
can be determined through
\begin{eqnarray}
V_{us} &=& \sin \theta_{12} \cos \theta_{13}, \quad
V_{ub} \; =\; \sin \theta_{13} \, e^{-i \delta_{13}} , \nn
V_{cb} & = & \sin \theta_{23}  \cos \theta_{13}  . \label{pdg}
\end{eqnarray}
The Wolfenstein parameterisation\cite{w} 
\begin{eqnarray}
\! V & \! = &
\left(
\begin{array}{@{}c@{~}c@{~}c@{}}
1 -\frac{{  \lambda^2} }{2} &{  \lambda} &
     \!\! A \lambda^3
     ({ {\rho}} \!-\! i { {\eta}} ) \\[2mm]
- \lambda
& 1 -  \frac{{  \lambda^2} }{2} &
 { A} {  \lambda^2} \\[2mm]
 A \lambda^3
(1 \!-\! \rho \!-\! i \eta  ) 
& - A \lambda^2 & 1
\end{array} \! \right) \; \label{w}
\end{eqnarray}
is an expansion of $V$ in terms of $\lambda \simeq 0.22$ to order
$\lambda^3$. It shows both the hierarchy of the CKM elements and their 
correlations, like $|V_{us}|\simeq |V_{cd}|$ and $|V_{cb}|\simeq |V_{ts}|$.
The apex of the standard unitarity triangle (UT), which is shown in 
\fig{fig:ut} is defined by\cite{blo}
\begin{eqnarray}
{ \ov{\rho}} + { i \ov{\eta}} & \equiv &
- \frac{{ V_{ub}^*} V_{ud}} {{V_{cb}^*} V_{cd}} 
\; =\; \lt| \frac{{ V_{ub}^*} V_{ud}}{{ V_{cb}^*} V_{cd}} \rt|
  e^{ i \gamma}
\label{ut}
\end{eqnarray}
$(\ov{\rho},\ov{\eta})$ coincide with $(\rho,\eta)$ up to corrections of
order $\lambda^2$. 
With \eq{ut} and 
\begin{eqnarray}
\lambda &\equiv& \sin \theta_{12}, \quad
A \lambda^2 \; \equiv \; \sin \theta_{23},
\label{wex}
\end{eqnarray}
the Wolfenstein parameterisation can be made exact\cite{blo}, that is
$V$ can be expressed in terms of $(\lambda, A, \ov{\rho}, \ov{\eta})$ to
any desired order in $\lambda$. In the following I always use the PDG
phase convention and the exact definitions in Eqs.~(\ref{ut}) and
(\ref{wex}), with one exception: I ignore the small phase of $-V_{cd}$
(see Ref.\cite{pdg}), so that I can identify $\arg V_{ub}^*=\delta_{13}$
with $\gamma$ and $\arg V_{td}^*$ with the angle $\beta$ of the
unitarity triangle. This approximation is correct to 0.1\%.

The numerical results presented in the following have been prepared with
the help of the \emph{Heavy Flavor Averaging Group (HFAG)}\cite{hfag}
and the \emph{CKMfitter}\cite{ckmf} and \emph{UTFit}\cite{utf} groups.   
\emph{CKMfitter} uses a Frequentist treatment of theoretical uncertainties,
while \emph{UTFit} pursues a Bayesian approach,  
using flat probability distribution functions for theoretical uncertainties.

\section{CKM elements from tree-level decays}\label{sect:tree}
\begin{figure}[t]
\epsfxsize0.5\columnwidth
\figurebox{}{}{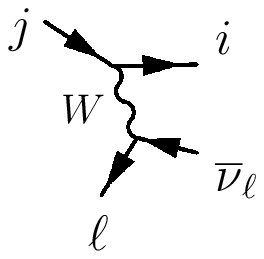}
\caption{$|V_{uj}|$ 
  and $|V_{cj}|$, $j=d,s,b$, are determined from semileptonic decays.}
\label{fig:sl}
\end{figure}
The standard way to determine the magnitudes of the elements of the
first two rows of $V$ uses semileptonic hadron decays, depicted in
\fig{fig:sl}. From \eq{w} one realizes that an accurate determination of
$V_{us}$ or $V_{ud}$ determines $V_{cd}$ and $V_{cs}$ as well. Therefore
measurements of semileptonic $c\to d$ and $c\to s$ decays are usually
viewed as test of the computation of the hadronic form factors entering 
the decay amplitudes. Charm decays are covered by
Iain Stewart.\cite{s} 

\subsection{$V_{ud}$}
$V_{ud}$ can be determined from superallowed ($0^+\to 0^+$) nuclear
$\beta$ decay and from the $\beta$ decays $n\to p\, \ell\, \ov{\nu}_\ell
(\gamma)$ and $\pi^- \to \pi^0\, \ell\, \ov{\nu}_\ell (\gamma)$. Since 
no other decay channels are open, the semileptonic decay rate can be 
accessed through lifetime measurements. All three methods involve the 
hadronic form factor of the vector current: 
\begin{eqnarray}
&& \bra{f} \ov{u} \gamma_\mu d \ket{i}, 
\no
\end{eqnarray}
where $(i,f)=(0^+,0^+),(n,p)$ or $(\pi^\pm,\pi^0)$. The neutron
$\beta$ decay further involves the {form factor} of the
{axial vector current}:
\begin{eqnarray}
\bra{f} \ov{u} \gamma_\mu\gamma_5 d \ket{i}
\no
\end{eqnarray}
The form factors parameterise the long-distance QCD effects, which bind
the quarks into hadrons. The normalisation of the vector current is
fixed at the kinematic point of zero momentum transfer $p_i-p_f$ in the
limit $m_u=m_d$ of exact isospin symmetry. The Ademollo-Gatto
theorem\cite{ag} assures that corrections are of second order in the
symmetry breaking parameter $(m_d-m_u)/\lhad $, where $\lhad$ is the
relevant hadronic scale. No such theorem protects the axial form factor
$\bra{p} \ov{u} \gamma_\mu\gamma_5 d \ket{n}$, but the corresponding
parameter $G_A$ can be extracted from asymmetries in the Dalitz plot.
Experimentally the highest precision in the determination of $V_{ud}$ is
achieved in the nuclear $\beta$ decay, but $n\to p\, \ell\,
\ov{\nu}_\ell (\gamma)$ starts to become competitive. However, there is
currently a disturbing discrepancy in the measurement of the neutron
lifetime among different experiments.\cite{h} From a theoretical point
of view progress in $n\to p\, \ell\, \ov{\nu}_\ell (\gamma)$ and,
ultimately, in the pristine $\pi^- \to \pi^0\, \ell\, \ov{\nu}_\ell
(\gamma)$ decay are highly desirable to avoid the nuclear effects of
$0^+ \to 0^+$ transitions. On the theory side QED radiative corrections
must be included to match the experimental accuracy, recently even
dominant two-loop corrections to $n\to p\, \ell\, \ov{\nu}_\ell
(\gamma)$ have been calculated.\cite{cms}

The world average for $V_{ud}$ reads\cite{c}:
\begin{eqnarray}
 V_{ud} &=& 0.9738 \pm 0.0005 \label{vud}
\end{eqnarray}

\subsection{$V_{us}$}\label{sect:vus}
\begin{figure}%
\epsfxsize\columnwidth
\figurebox{}{}{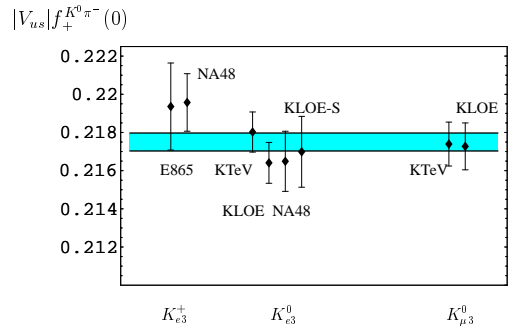}
\caption{$V_{us} f_+^{K^0\pi^-} (0)$. The horizontal band is the range 
quoted in \eq{fvusres}. Courtesy of Vincenzo Cirigliano.\protect\cite{c}}
\label{fig:vus}
\end{figure}

$V_{us}$ can be determined from Kaon and $\tau$ decays. The most
established method uses the so-called $K\ell 3$ decays $ K^0 \to \pi^-
\ell^+ \nu_\ell$, $ K^0 \to \pi^- \mu^+ \nu_\ell$, $ K^+ \to \pi^0
\ell^+ \nu_\ell$ and $ K^+ \to \pi^0 \mu^+ \nu_\ell$. 
The decay rates schematically read
\begin{eqnarray}
\lefteqn{ \Gamma (K\to \pi \ell^+\nu_\ell)
  \; \propto } \nn 
&& { V_{us}^2} \, { \lt| f_+^{K^0\pi^-} (0)
          \rt|^2} 
\lt[ 1 \; +\; 2 { \Delta_{SU(2)}^K} \; +\;
                         2 {  \Delta_{\rm em}^{K\ell}} \rt] .\no
\end{eqnarray}
The hadronic physics is contained in 
\begin{eqnarray}
\lefteqn{
\bra{\pi^-(p_\pi)} \ov{s}\gamma^\mu u \ket{K^0(p_K)} \; =}\nn
&& f_+^{K^0\pi^-} (0) (p_K^\mu+p_\pi^\mu ) + 
     {\cal O} (p_K-p_\pi) \nn
&& \!\! \Delta_{SU(2)}^{K^+} \; =\;
    \frac{ f_+^{K^+\pi^0} (0)}{ f_+^{K^0\pi^-} (0)}-1, \quad
   \Delta_{SU(2)}^{K^0} \; =\; 0. \no
\end{eqnarray}
and QED corrections are contained in $\Delta_{\rm em}^{K\ell}$. The
Ademollo-Gatto theorem\cite{ag} ensures $f_+^{K^0\pi^-} (0) = 1 + {\cal
  O}( (m_s-m_d)^2/\Lambda_{\rm had}^2)$.  $f_+^{K^0\pi^-} (0) - 1$ can
be calculated with the help of Chiral Perturbation Theory
($\chi$PT)\cite{gl}, which exploits the fact that the pseudoscalar
mesons are Goldstone bosons of a dynamically broken chiral symmetry of
QCD. $\chi$PT amounts to a systematic expansion in $ p/\Lambda_{\rm
  had}$, $ M/\Lambda_{\rm had}$, $ m_\ell/\Lambda_{\rm had}$ and the
electroweak coupling $ e$.  Here $ p$ and $ M$ denote meson momenta and
masses and $m_\ell$ is the lepton mass. There has been a substantial
progress in the calculation of both $\Delta_{\rm em}^{K\ell}$\cite{em}
and $f_+^{K^0\pi^-} (0)$\cite{ff} since \emph{LP'03}.  Significant
effects of ${\cal O} (e^2 p^2) $ QED corrections on differential
distributions were found; they must be included in Monte Carlo
simulations. The value for $V_{us} f_+^{K^0\pi^-} (0)$ extracted from
various experiments is shown in \fig{fig:vus}.
The world average reads:\cite{c}
\begin{eqnarray}
f_+^{K^0\pi^-} V_{us} & = & 0.2175\pm 0.0008 .
\label{fvusres}
\end{eqnarray}
Combining the results from $\chi PT$ at order $p^6$ and quenched lattice gauge
theory (new) to\cite{ff}
\begin{eqnarray}
f_+^{K^0\pi^-}=0.972\pm 0.012
\no
\end{eqnarray}
one arrives at 
\begin{eqnarray}
 V_{us}=0.2238\pm 0.0029
\label{vusres3}
\end{eqnarray}
from $K\ell3$.

$V_{us}$ can also be determined from the $K\mu 2$ decay $ K^+ \to \mu^+
\nu_\mu(\gamma)$.\cite{m} The hadronic quantity entering this decay
is the Kaon decay constant $F_K$. Uncertainties can be better controlled
in the ratio $F_K/F_\pi$ and one considers
\begin{eqnarray}
\lefteqn{\frac{\Gamma (K^+ \to \mu^+ \nu_\mu(\gamma))}{
 \Gamma (\pi^+ \to \mu^+ \nu_\mu (\gamma))}
 \; = } \nn
&& \frac{ V_{us}^2}{V_{ud}^2} \,
          { \frac{F_K^2}{F_\pi^2} }\,
          \frac{M_K^2-m_\mu^2}{M_\pi^2-m_\mu^2}\,
           \, \lt[ 1  -  { \frac{\alpha}{\pi} \lt(C_\pi-C_K\rt) }
                  \rt]
\no
\end{eqnarray}
with QED corrections $C_\pi-C_K\rd =3.0\pm 1.5$. Using the
result\cite{milc} $F_K/F_\pi =1.210 \pm 0.004\pm 0.013$ computed by the
MILC collaboration with 2+1 dynamical quarks, one finds (with $V_{ud}$
from \eq{vud}): 
\begin{eqnarray}
V_{us}=0.2223\pm 0.0026 \label{vusres2}
\end{eqnarray}
from the $\rd K\mu 2$ decay. This is astonishingly precise and $ K^+ \to
\mu^+\nu_\mu(\gamma)$ may constrain mass and couplings of a charged
Higgs boson, which can mediate this decay as well.\cite{m} 

\begin{figure}
\epsfxsize0.7\columnwidth
\figurebox{}{}{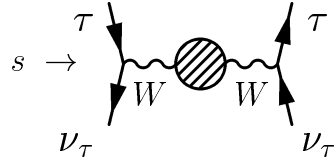}
\caption{The optical theorem relates $\Gamma^{\Delta S=0,1} 
  (\tau \to \mbox{hadrons } \nu_\tau(\gamma))$ to $\Pi^{T,L}_{s,d}$. The
  blob denotes the hadronic states contributing to $\Pi^{T,L}_{s,d}$.
  The leading term in the OPE is obtained by replacing the blob by a
  $(u,d)$ or $(u,s)$ quark loop and gluons to the desired order in
  $\alpha_s$.}
\label{fig:tope}
\end{figure}

The third possibility to measure $V_{us}$ used hadronic $\tau$ decays to 
the inclusive final state with strangeness $|S|=1$. The experimental
inputs are the ratios
\begin{eqnarray}
\rd R_{\tau s} &\rd =& \rd \frac{\Gamma^{\Delta S=1} (\tau \to
       \mbox{hadrons } \nu_\tau(\gamma))}{\Gamma (\tau \to e \ov{\nu}_e
         \nu_\tau(\gamma))}
\; \propto \; \gn V_{us}^2 \nn
\rd R_{\tau d} &\rd =& \rd \frac{\Gamma^{\Delta S=0} (\tau \to
       \mbox{hadrons } \nu_\tau(\gamma))}{\Gamma (\tau \to e \ov{\nu}_e
         \nu_\tau(\gamma))}
\; \propto \; \gn V_{ud}^2 \no
\end{eqnarray}
Here $\rd S$ is the strangeness. The optical theorem allows to relate
$\rd R_{\tau s,d}$ to the {\rd QCD current-current
  correlators} $\gn \Pi^T_{s,d}$ and $\gn \Pi^L_{s,d}$:
\begin{eqnarray}
\lefteqn{\!\! R_{\tau s,d} \; = \; \rd
12 \pi\int_0^1 \! d\,z (1-z)^2 \times} \nn
&&      \quad        \, \lt[ (1 + 2 z) \, {\gn  \imag \Pi^T_{s,d} (z)} \; +\;
                  {\gn \imag \Pi^L_{s,d} (z) } \rt]   \no
\end{eqnarray}
with $z=s/M_\tau^2=(p_\tau-p_{\nu_\tau})^2/M_\tau^2$. This relationship
is depicted in \fig{fig:tope}.
$\Pi^{T,L}_{s,d}$ can be computed through an {\rd operator product
  expansion (OPE)}. The leading term is {\rd massless perturbative QCD},
subleading operators entering $\Pi^{T,L}_s$ are $\rd m_s^2$ and $\rd m_s
\langle \ov{q} q \rangle$. The OPE amounts to an expansion in
$\lqcd/M_\tau$, $m_s/M_\tau$ and $\alpha_s(m_\tau)$.  In the limit
$m_s=0$ of exact SU(3)$_{\rm F}$ symmetry the ratio $R_{\tau s}/R_{\tau
  d}$ would directly determine $V_{us}^2/V_{ud}^2$. Hence it suffices to
compute the (small) SU(3)$_{\rm F}$ breaking quantity\cite{pp}
\begin{displaymath}
\delta R_\tau \;\equiv\;
\frac{R_{\tau d}}{V_{ud}^2} \,-\,\frac{R_{\tau s}}{V_{us}^2}. \no
\end{displaymath}
With $\delta R_\tau \;=\; 0.218 \,\pm\, 0.026$\cite{gjpps} 
and experimental data from OPAL\cite{opal} one finds 
$\rd R_{\tau d}=3.469 \pm 0.014$,
$\rd R_{\tau s}=0.1694 \pm 0.0049$ and finally:\cite{gjpps} 
\begin{eqnarray}
 V_{us} &= & \sqrt{\frac{R_{\tau s}}
{R_{\tau d}/|V_{\!ud}|^2\,-\,\delta R_\tau}}\nn
& = & 0.2219 \pm 0.0033_{\rm exp} \pm 0.0009_{\rm th}\nn
& = & 0.2219 \pm 0.0034. \label{vusrest}
\end{eqnarray}
The dominant source of uncertainty in $\delta R_\tau$, which enters
\eq{vusrest} as a small correction, is from $m_s$. In the near future it
should be possible to improve on $V_{us}$ with data from BaBar and
BELLE. 

In summary one finds an excellent consistency of the three numbers for
$V_{us}$ from $K\ell3$, $\rd K\mu 2$ and $\tau$ decays. This is
remarkable, since the three methods use very different theoretical tools
to address the strong interaction: Chiral perturbation theory, lattice
gauge theory and the operator product expansion. The result nicely
reflects the tremendous progress of our understanding of QCD at low
energies. Averaging the results of Eqs.~(\ref{vusres3}), (\ref{vusres2})
and (\ref{vusrest}) one finds:
\begin{eqnarray}
   V_{us} & = & 0.2227\pm 0.0017 \label{vus}
\end{eqnarray}
With $V_{ud}$ in \eq{vud} one can perform the first-row unitarity check 
\begin{eqnarray}
   V_{us}^2 + V_{ud}^2 + |V_{ub}|^2 -1 & \simeq &
   V_{us}^2 + V_{ud}^2 -1 \nn 
& = & -0.0021  \pm 0.0012 \no
\end{eqnarray}
The {\gn Cabibbo matrix} is unitary at the $\gn 1.8\sigma$ level, 
just as at \emph{LP'03}:\cite{sch}
\begin{eqnarray}
   V_{us}^2 + V_{ud}^2 -1 \; =\; -0.0031 \pm 0.0017 \no
\end{eqnarray}

\subsection{$V_{cb}$}
$V_{cb}$ can be determined from inclusive or exclusive $b\to c\ell
\nu_\ell$ decays. Exclusive decays 
are not discussed here. The analysis of the inclusive decay employs
an OPE\cite{sv}, similarly to the determination of $V_{us}$ from $\tau$ decay
discussed in Sect.~\ref{sect:vus}.
\begin{figure}
\epsfxsize\columnwidth
\figurebox{}{}{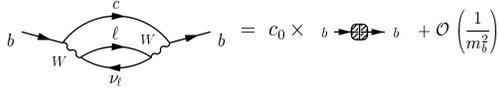}
\caption{OPE for $\ov{B}\to X_c \ell \ov{\nu}_\ell$. The leading 
operator $\ov{b}b$ has dimension 3.}\label{fig:vcbope}
\end{figure}
The optical theorem relates the inclusive decay rate $\ov{B}\to X_c \ell
\ov{\nu}_\ell$ to the imaginary part of the $B$ meson self energy,
depicted on the LHS of \fig{fig:vcbope}. The OPE matches the self energy 
diagram to matrix element of effective operators, whose coefficients 
contain the short-distance information associated with the scale $m_b$ 
and can be calculated perturbatively. Increasing dimensions of the
operators on the RHS of \fig{fig:vcbope} correspond to decreasing powers
of $m_b$ in the coefficient functions, so that the OPE amounts to a
simultaneous expansion in $\lqcd/m_b$ and $\alpha_s (m_b)$. Since 
$\bra{B} \ov{b} b\ket{B}=1+{\cal O}(\lqcd^2/m_b^2)$ and there are no
dimension-4 operators, non-perturbative parameters first
occur at order $\lqcd^2/m_b^2$. They are 
\begin{eqnarray}
{\gn \mu_\pi^2} &\propto&  -\bra{B} \ov{b} D_\perp^2 b \ket{B}\nn
{\gn \mu_G^2} &\propto& \bra{B} \ov{b} i \sigma_{\mu\nu}G^{\mu\nu} b \ket{B}
     \no
\end{eqnarray}
$ \mu_G^2$, which parameterises the matrix element of the chromomagnetic
operator, can be determined from spectroscopy.
Hence to order $\rd \lqcd^2/m_b^2$ one only has to deal with the three 
quantities $\rd m_b$, $\rd m_c$  and $\rd \mu_\pi^2$, which quantifies
the Fermi motion of the $b$ quark inside the B meson. 

\begin{figure}
\epsfxsize\columnwidth
\figurebox{}{}{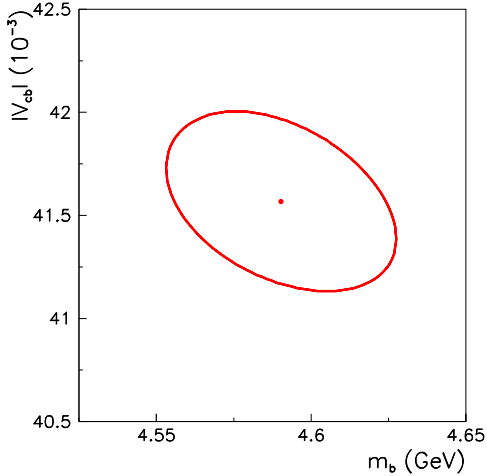}
\caption{Fit result for $V_{cb}$ vs.\ $m_b$, which is defined in the
  kinetic scheme.\protect\cite{gu} Fit and plot are courtesy of Oliver
  Buchm\"uller and Henning Fl\"acher. See also\protect\cite{bf}.}
\label{fig:vcb}
\end{figure}

The {\rd OPE} can further be applied to certain {\rd spectral moments}
of the $\rd B\to X \ell\ov{\nu}_\ell$ decay, the distributions of the
{\rd hadron invariant mass} $\rd M_X$ and of the {\rd lepton energy}.
Further the same parameters govern {\rd different} inclusive decays, for
instance also $B\to X_s \gamma$.  Therefore there is a lot of redundancy
in the determination of $V_{cb}$, providing powerful checks of the
theoretical framework.  The state of the art are fits to order $\rd
\lqcd^3/m_b^3$, which involve {\rd 7 parameters}.\cite{gu} The result of
a global fit to {\rd hadron} and {\rd lepton moments} in $\gn B\to
X\ell\nu_{\ell}$ and {\rd photon energy moments} in $\gn B\to X_s\gamma$
from {\gn BaBar, BELLE, CDF, CLEO, DELPHI}\cite{bf} can be seen in
\fig{fig:vcb}.
It gives
\begin{eqnarray}
V_{cb} &=& 41.6 \pm 0.3_{\rm exp}
       \pm 0.3_{\rm OPE\,moments} \nn 
  && \phantom{41.6} \pm 0.3_{\rm OPE\,\Gamma_{sl}}\nn
&=& (41.6 \pm 0.5) \cdot 10^{-3} \label{vcb}
\end{eqnarray}
from inclusive $B\to X\ell\nu_{\ell}$.

\subsection{$|V_{ub}|$}
I discuss the determination of $|V_{ub}|$ from inclusive $B\to X_u\ell
\nu_\ell$ decays. Exclusive decays are discussed in.\cite{s} In
principle one could determine $|V_{ub}|$ in the same way as $V_{cb}$, if
there were no background from $B\to X_c\ell \nu_\ell$ decays. Its
suppression forces us to impose cuts on the lepton energy $E_\ell$, the
hadronic energy $E_X$, the hadron invariant mass $M_X$ or a judiciously
combination of them.  $M_X$ is too small for an OPE in the portion of
phase space passing these cuts.  Still some components of the hadron
momentum $\vec P_X$ are large.  The description of inclusive $B$ decays
in this region involves the non-perturbative {shape function $S$}, which
is a parton distribution function of the {B meson}.  At leading order in
$1/m_b$ the {same} $S$ governs the photon spectrum in $B\to X_s \gamma$
and differential decay rates in $B\to X_u \ell\ov{\nu}_\ell $. This
allows us to extract $S$ from $B\to X_s \gamma$ for the use in $\rd B\to
X_u \ell\ov{\nu}_\ell $. The goal to reduce the theoretical uncertainty
below 10\% requires to understand corrections in both expansion
parameters $\alpha_s$ and $\lqcd/m_b$. For a correct treatment of
radiative QCD corrections one must properly relate the differential
decay rate $d\Gamma$ to the shape function $S$. This is achieved by a
factorisation formula, which has the schematic form:\cite{bm}
\begin{eqnarray}
d\Gamma & \propto & H \, \int_0^{P_+} \! d\omega \;
            J\lt( m_b(P_+ -\omega) \rt) \,
                S(\omega) \no
\end{eqnarray}
Here $H$ contains the hard QCD, associated with scales of order $m_b$.
The jet function $J$ and the shape function $S$ contain the physics from
scales of orders $M_X\sim \sqrt{m_b\lqcd}$ and $\lqcd$, respectively.
$P_+$ and $P_-$ are defined as $P_\pm =E_X\mp |\vec{P}_X|$. From $\lqcd
\ll P_+ \sim M_X \sim \sqrt{m_b\lqcd} \ll P_- \leq m_b $ one realizes
that one has to deal with a multi-scale problem, which is more
complicated than $B\to X_c\ell \nu_\ell$. The second frontier of
research in $B\to X_u\ell \nu_\ell$ deals with subleading shape
functions $s_i$, which occur at order $1/m_b$. They are different in
$B\to X_u\ell \nu_\ell$ and $B\to X_s \gamma$, but their moments can be
related to OPE parameters like $\mu_\pi^2$, which gives some guidance to
model these functions.\cite{ls} Meanwhile an event generator for $B\to
X_u \ell\ov{\nu}_\ell $ decays is available,\cite{lnp} with formulae
which contain all available theoretical information and smoothly
interpolate between the shape function and OPE regions.  It is pointed
out that a cut on the variable $P_+$, which is directly related to the
photon energy in $B\to X_s \gamma$, makes the most efficient use of the
$S(\omega)$ extracted from the radiative decay.\cite{mr,lnp}
Alternatively one can eliminate $S(\omega)$ altogether by forming proper
weighted ratios of the endpoint photon and lepton spectra in $B\to X_s
\gamma$ and $B\to X_u\ell \nu_\ell$, respectively.\cite{lr} Using also the 
information from $B\to X_c\ell \nu_\ell$ on $m_b$ and $\mu_\pi^2$ the
data from CLEO\cite{cvub}, BELLE\cite{bevub} and BaBar\cite{bavub}
combine to the world average\cite{hfag} 
\begin{eqnarray}
V_{ub} &=& (4.39 \pm 0.20_{\rm exp}
       \pm 0.27_{\rm th,m_b,\mu_\pi^2}) \cdot 10^{-3} \nn 
&=& (4.39 \pm 0.34) \cdot 10^{-3} \label{vub}
\end{eqnarray}
from inclusive $B\to X_u\ell\nu_{\ell}$. \eq{ut} implies that
  $|V_{ub}/V_{cb}|$ defines a circle in the $(\ov{\rho},\ov{\eta})$
  plane which is centered around $(0,0)$. With Eqs.~(\ref{vcb}) and
  (\ref{vub}) its radius is constrained to 
\begin{eqnarray}
 R_u \; \equiv \; {\gn \sqrt{\ov{\rho}^2+\ov{\eta}^2}} &=& 0.45 \pm 0.04 .
   \label{ru}
\end{eqnarray}

\subsection{$\rd \arg V_{ub}$}
\begin{figure}
\epsfxsize\columnwidth
\figurebox{}{}{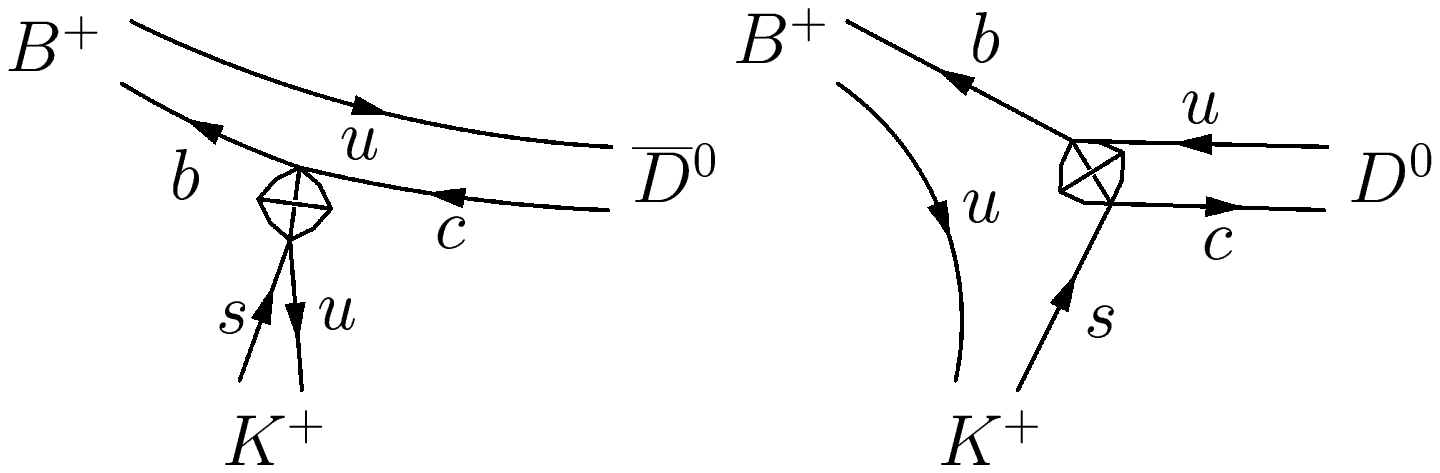}
\caption{The Gronau-London-Wyler method combines the rates of 
$\protect B^\pm \to \protect \DorDbar\,[\to f_i] K^\pm $
for different final states \protect $f_i$.}\label{fig:glw}
\end{figure}

$\gamma= \arg V_{ub}^*$ can be determined from exclusive $\rd B \to
\DorDbar\, X$ decays, where $X$ denotes one or several charmless mesons.
This method exploits the interference of the tree-level $\rd b\to
c\ov{u}q$ and $\rd b\to u\ov{c}q$ amplitudes, where $q=d$ or $q=s$.
The prototype is the Gronau-London-Wyler (GLW) method\cite{glw} shown in 
\fig{fig:glw}. 
The decays $\rd B \to D^0 X$ and $\rd B \to \Dbar X$ interfere, if both
subsequent decays $\rd D^0\to f$ and $\rd \Dbar \to f$ are allowed.  One
needs four measurements to solve for the magnitudes of the $b\to c$ and
$b\to u$, their relative strong phase and their relative weak phase,
which is the desired UT angle $\gamma$. For example one can combine the
information of the branching fractions of $\rd B^+\to \DorDbar\, [\to
K^\pm \pi^\mp] K^+$ and $\rd B^\pm \to \DorDbar\, [\to \pi^+ \pi^-]
K^\pm$. This works with untagged non-flavour-specific decays as
well:\cite{ggssz} E.g.\ the final state $\DorDbar\, \phi$ does not
reveal whether the decaying meson was a $B_s$ or $\Bbar_s$. Still, when
at least three pairs of $\rd \BsorBsbar \to \DorDbar\, [\to f_i ] \phi $
and $\rd \BsorBsbar \to \DorDbar\, [\to \ov{f}_i ] \phi $ branching
fractions are measured, where $\rd \ov{f}_i=C\!P f_i $ (and the $\rd
f_i$'s are not CP eigenstates), one has enough information to solve for
$\gamma$. Since no flavour tagging is involved, the Tevatron experiments
may contribute to these class of $\gamma$ determinations. The described
determination of $\gamma$ from tree-tree interference is modular, that
is measurements in different decay modes can be combined, as they partly
involve the same hadronic parameters. One should further first average
the branching ratios from different experiments and then determine
$\gamma$ instead of averaging the inferred values of $\gamma$ obtained
from different experiments.  Combining (almost) all $\rd B^+ \to D^0
K^{+(*)}$ data gives (preliminary)\cite{ckmf}
\begin{eqnarray}
{\gn \gamma} &=& (70  \epm{12}{14})^\circ \label{gad}
\end{eqnarray}
and the second solution $\gamma-180^\circ \sim - 110^\circ$.
This is $\gn \gamma\rd =\arg V_{ub}^*$ determined from the {\rd tree-level}
$\rd b\to u\ov{c} s$ amplitude.

Within the Standard Model $b\to u\ov{u} d$ decays of tagged $B^0$ mesons
are used to determine the UT angle $\alpha$. $b\to u\ov{u} d$ decays
involve both a tree and a penguin amplitude.  The penguin component can
be eliminated, if several decay modes related by isospin are combined,
as in the Gronau-London method\cite{glo} which uses $B^+\to \pi^+ \pi^0$,
$B^0\to \pi^+ \pi^-$ and $B^0\to \pi^0 \pi^0$. The $B\to \rho\pi$ and
$B\to \rho\rho$ decay modes are better suited for the determination of
$\alpha$, because the penguin amplitude is smaller. A combined analysis 
of the $\pi\pi$, $\rho\pi$ and $\rho\rho$ systems gives 
\begin{eqnarray}
\rd \alpha_{\rm exp} 
  &\rd =& \rd (99\epm{12}{9})^\circ  \label{alpha}
\end{eqnarray}
and the second solution $\alpha_{\rm exp}\!-\! 180^\circ \sim -81^\circ$. 
The experimental result $\alpha_{\rm \exp}$ could differ from the true 
$\alpha=\arg(-V_{tb}^*V_{td}/(V_{ub}^*V_{ud}))$, if new physics alters
the \bbmd\ amplitude. However, the influence from new physics is fully
correlated in $\alpha_{\rm exp}$ and the CP asymmetry measured in 
$b\to c \ov{c} s$ decays. From the latter (see \eq{beta} below) we infer 
the \bbmd\ phase $2\beta_{\rm exp}=(43.7 \pm 2.4)^\circ $. The  \bbmd\
phase cancels from the combination 
$2\gamma=360^\circ - 2 \alpha_{\rm exp}- 2\beta_{\rm exp}$, so that one
obtains
\begin{eqnarray}
\gn \gamma &\rd =&\rd (59 \epm{9}{12})^\circ \label{gaa}
\end{eqnarray}
and the second solution $\gamma-180^\circ \sim - 121^\circ$.  Since the
isospin analysis eliminates the penguin component, this is $\gn
\gamma\rd =\arg V_{ub}^*$ determined from the {\rd tree-level} $\rd b\to
u\ov{u} d$ amplitude.

\begin{figure}
\epsfxsize\columnwidth
\figurebox{}{}{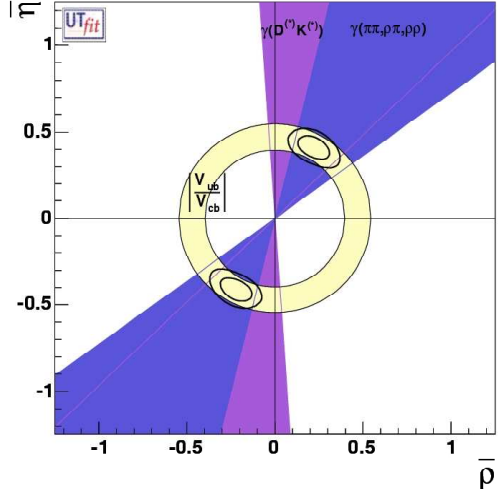}
\caption{UT from tree quantities alone. The annulus is the constraint
  in \eq{ru} derived from $|V_{ub}|$. The dark shadings correspond to 
  $\gamma$ from Eqs.~(\protect\ref{gad}) and (\protect\ref{gaa}).
  Courtesy of Maurizio Pierini.}\label{fig:uttree}
\end{figure}

The results in Eqs.~(\ref{gad}) and (\ref{gaa}) are in good agreement.
Their naive average is 
\begin{eqnarray}
\gn \gamma &\rd =&\rd (63 \epm{7}{9})^\circ . \label{gatree}
\end{eqnarray}
The successful determination of a CP phase from a tree-level amplitude is a 
true novel result compared to \emph{LP'03}. For the first time we can 
determine the UT from tree-level quantities alone, the result is
shown in \fig{fig:uttree}.
This is important, because the tree-level UT can only be mildly affected
by new physics and therefore likely determines the true values of 
$\ov{\rho}$ and $\ov{\eta}$.

\section{CKM elements from FCNC processes}\label{sect:loop}
In the Standard Model FCNC processes are suppressed by several effects:
First they only proceed through electroweak loops. Second they come with
small CKM factors like $|V_{ts}|\sim 0.04$ and $|V_{td}|\sim 0.01$.
Loops with an internal charm quark are further suppressed by a factor of
$m_q^2/M_W^2$ from the GIM mechanism. Radiative and leptonic decays
further suffer from an additional helicity suppression, because only
left-handed quarks couple to W bosons and undergo FCNC transitions.  All
these suppression mechanism are accidental, resulting from the particle
content of the Standard Model and the unexplained smallness of most
Yukawa couplings. They are absent in generic extensions of the Standard
Model (like its supersymmetric generalisations) making FCNC highly
sensitive to new physics, probing scales in the range of 200\gev\ to
100\tev, depending on the model considered. This feature is a major
motivation for the currently performed high-statistics experiments in
flavour physics. Comparing different constraints on the UT from FCNCs
processes and the tree-level constraints discussed in
Sect.~\ref{sect:tree} therefore provides a very powerful test of the
Standard Model. 

\subsection{Meson-antimeson mixing}
\begin{figure}
\epsfxsize0.7\columnwidth
\figurebox{}{}{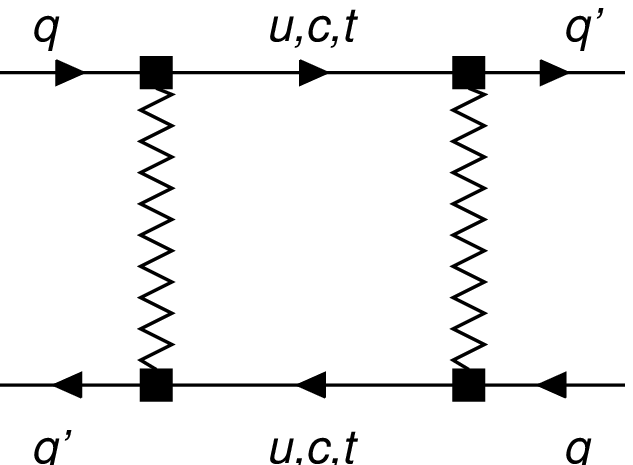}
\caption{Meson-antimeson mixing. $(q,q^\prime)=(s,d)$, $(b,d)$ and
  $(b,s)$ for \kkm, \bbmd\ and \bbms, respectively.}
\label{fig:box}
\end{figure}

\kkm, \bbmd\ and \bbms\ are all induced by box diagrams, depicted in 
\fig{fig:box}.
Each meson-antimeson system involves two mass eigenstates, 
their mass difference \dm\ measures the magnitude of the box diagram and
therefore constrains magnitudes of CKM elements. The phase of box
diagram and thereby the phases of the CKM elements involved are
constrained through CP-violating quantities. Tab.~\ref{tab:mix} shows the
relationship of the measurements to the CKM phenomenology. 
\begin{table*}[tb]
\begin{tabular}{l|ccc}
 & \kkm & \bbmd & \bbms \\\hline
CP-conserving  
quantity: & $\dm_K$ &    $\dm_{B_d}$ &     $\dm_{B_s}$ \\
CKM information: & $|V_{cs} V_{cd}|^2$ & $|V_{tb} V_{td}|^2$ &
                 $|V_{tb} V_{ts}|^2$ \\ 
UT constraint: & none & $R_t =\sqrt{(1-\ov{\rho})^2+\ov{\eta}^2}$ & 
               none \\\hline
CP-violating quantity: 
  & $\epsilon_K$ & $ a_{\rm CP}^{\rm mix} (B_d\to J/\psi K_S)$ & 
            $ a_{\rm CP}^{\rm mix} (B_s\to J/\psi \phi) $ \\
CKM information: & $\imag (V_{ts}V_{td}^*)^2$ & 
               $\ds \sin(2\beta)$ &
             $\ds \sin(2\beta_s)$ \\ 
UT constraint: & $\ov{\eta}[ (1-\ov{\rho}) + \mbox{\rm const.}]$
& $\ds \frac{\ov{\eta}}{1-\ov{\rho}}$ & $\ds \ov{\eta}$ \\ 
\end{tabular}   
\caption{Relationship of meson-antimeson mixing to CKM and UT
  parameters. $\ds \beta = \arg V_{td}^*$ is one of the UT angles in
  \fig{fig:ut} and 
  $\beta_s = \arg (-V_{ts})\simeq \lambda^2 \ov{\eta}$. 
}\label{tab:mix}
\end{table*}
The quantities in the first two columns of Tab.~\ref{tab:mix} are
well-measured and there is a lower bound on $\dm_{B_s}$. 

\subsection{$\epsilon_K$}
While $\epsilon_K$, which quantifies indirect CP violation in $K\to \pi
\pi$ decays, is measured at the percent level, its relationship to $\imag
V_{td}^{*2}\propto \ov{\eta} (1-\ov{\rho})$ is clouded by hadronic
uncertainties in the matrix element
\begin{eqnarray}
\bra{K^0} \ov{d}s_{V-A} \ov{d}s_{V-A} \ket{\ov{K}{}^0} &\equiv& 
  \frac{8}{3} \, f_K^2 M_K^2\,  B_K .\no
\end{eqnarray}
This defines the hadronic parameter $B_K$, which must be
computed by non-perturbative methods like lattice QCD. $M_K$ and $f_K$
are the well-known mass and decay constant of the Kaon. This field has
experienced a major breakthrough since \emph{LP'03}, since meanwhile
fully unquenched computations with 2+1 dynamical staggered quarks are
available. Using MILC configurations the HPQCD collaboration reports a
new result\cite{hpqcdbk}\footnote{In my talk I reported the preliminary
  value $B_K=0.630\pm 0.018_{\rm stat}\pm 0.015_{\rm ch.\ extr.} \pm
  0.030_{\rm disc.} \pm 0.130_{\rm p.\ match.}$.}
\begin{eqnarray}
\lefteqn{B_K (\mu=2\gev) \; =} \nn 
&&       0.618 \pm 0.018_{\rm stat} \pm 0.019_{\rm chiral\; extrapolation} \nn
&& \phantom{0.630}
  \pm 0.030_{\rm discret.} \pm 0.130_{\rm pert.\ matching} \nn
&=&0.618 \pm 0.136
\label{bk}
\end{eqnarray}
in the $\ov{\rm MS}$--NDR scheme. The conventionally used renormalisation
scale and scheme independent parameter reads
\begin{eqnarray}
\widehat{B}_K 
&=&0.83 \pm 0.18 \label{bkh}
\end{eqnarray}
The uncertainty from the perturbative lattice--continuum matching
dominates over the statistical error and the errors from chiral
extrapolation and discretisation in \eq{bk}. This matching calculation was
performed in\cite{bgm}. The error in \eq{bk} is a conservative estimate
of the unknown two-loop contributions to this matching. If one instead
takes twice the square of the one-loop result of\cite{bgm} as an
estimate of the uncertainty, one finds $0.036$ instead of $0.130$ in
\eq{bk} and
\begin{eqnarray}
B_K  (\mu=2\gev) &=&  0.618 \pm 0.054 \nn
\widehat{B}_K &=&0.83 \pm 0.07 \label{bkh2}
\end{eqnarray}
$\epsilon_K$ fixes $\ov{\eta} (1-\ov{\rho})$, so that it defines a hyperbola
in the $(\ov{\rho},\ov{\eta})$ plane. 

\subsection{$V_{td}$ from \bbmd}
The \bbmd\ mixing amplitude involves the hadronic matrix element 
\begin{eqnarray}
\rd \bra{B^0} \ov{b}d_{V-A} \ov{b}d_{V-A} \ket{\ov{B}{}^0} &=&
\frac{8}{3} \, M_{B_d}^2 \, f_{B_d}^2 B_{B_d}. \no
\end{eqnarray}
Since the decay constant $f_{B_d}$ is not measured, the whole
combination $f_{B_d}^2 B_{B_d}$ must be obtained from lattice QCD.  The
hadronic matrix element, however, cancels from the ``gold-plated''
mixing induced CP asymmetry $a_{\rm CP}^{\rm mix} (B_d\to J/\psi K_S)$,
which determines $\beta = \arg V_{td}^*$ essentially without hadronic
uncertainties. Combining all data from $b\to c\ov{c} s$ modes results
in\cite{twb,hfag} 
\begin{eqnarray}
    \sin (2\beta) \;= &&\!\!\!\! 0.69 \pm 0.03, \quad \cos(2\beta) >0  \nn
{\gn \Rightarrow}\quad \arg \lt( \pm V_{td}^*\rt) &=& \beta
  \; = \; (21.8 \pm 1.2)^\circ . \label{beta}
\end{eqnarray}
 
The precisely measured $\dm_{B_d} = 0.509 \pm 0.004\, {\rm ps}^{-1}$ is
proportional to $|V_{td}|^2 f_{B_d}^2 B_{B_d}$.  The HPQCD collaboration
has computed $f_{B_d}=216\pm 22\mev$ with 2+1 dynamical staggered
quarks.\cite{hpqcdfb} This measurement is discussed in detail
in\cite{s}. Combining this with $B_{B_d}$ from older quenched
calculations results in $f_{B_d} \sqrt{\widehat{B}_{B_d}} = (246 \pm
27)\mev$, where $\widehat{B}_{B_d}=1.52 B_{B_d}(\mu=m_b) $ is the
conventionally used scale and scheme independent variant of $B_{B_d}$.
Then from $\dm_{B_d}$ alone we find
\begin{eqnarray}
|V_{td}| = 0.0072 \pm 0.0008, \no
\end{eqnarray}
where the error is reduced by a factor of 2/3 compared to the old
determination from quenched lattice QCD.  

\subsection{$\rd |V_{td}|/|V_{ts}|$ from \bbm}
A measurement of the {\rd ratio $\dm_{B_d}/\dm_{B_s}$} will determine
$\rd |V_{td}|/|V_{ts}|$ via
\begin{eqnarray}
\lt| \frac{V_{td}}{V_{ts}} \rt| &=&
   \sqrt{\frac{\dm_{B_d}}{\dm_{B_s}}} \,
   \sqrt{\frac{M_{B_s}}{M_{B_d}}} \, {\gn \xi} \no
\end{eqnarray}
with the hadronic quantity $\xi = f_{B_s} \sqrt{\widehat{B}_{B_s}}/(
f_{B_d} \sqrt{\widehat{B}_{B_d}})$ which equals $\rd \xi=1$ in the limit
of exact {\rd SU(3)$_{\rm F}$}.  A new unquenched HPQCD result for
$f_{B_s}/f_{B_d}$\cite{hpqcdfb} presented in\cite{s} can be used to
refine the prediction for $\xi$.  The lower bound {\rd $\dm_{B_s}\geq
  14.5\,$ps$^{-1}$} implies $|V_{td}/V_{ts}|\leq 0.235$
which constrains one side of the unitarity triangle:
\begin{eqnarray}
\gn R_t \; \equiv \; {\gn \sqrt{(1-\ov{\rho})^2 +\ov{\eta}^2 }} &=&
    \lt| \frac{V_{td}}{V_{ts}\lambda} \rt|
   \; \leq \; 1.06  \no
\end{eqnarray}
 
\subsection{Global fit to the unitarity triangle}
\begin{figure*}
\epsfxsize1.1\columnwidth
\figurebox{}{}{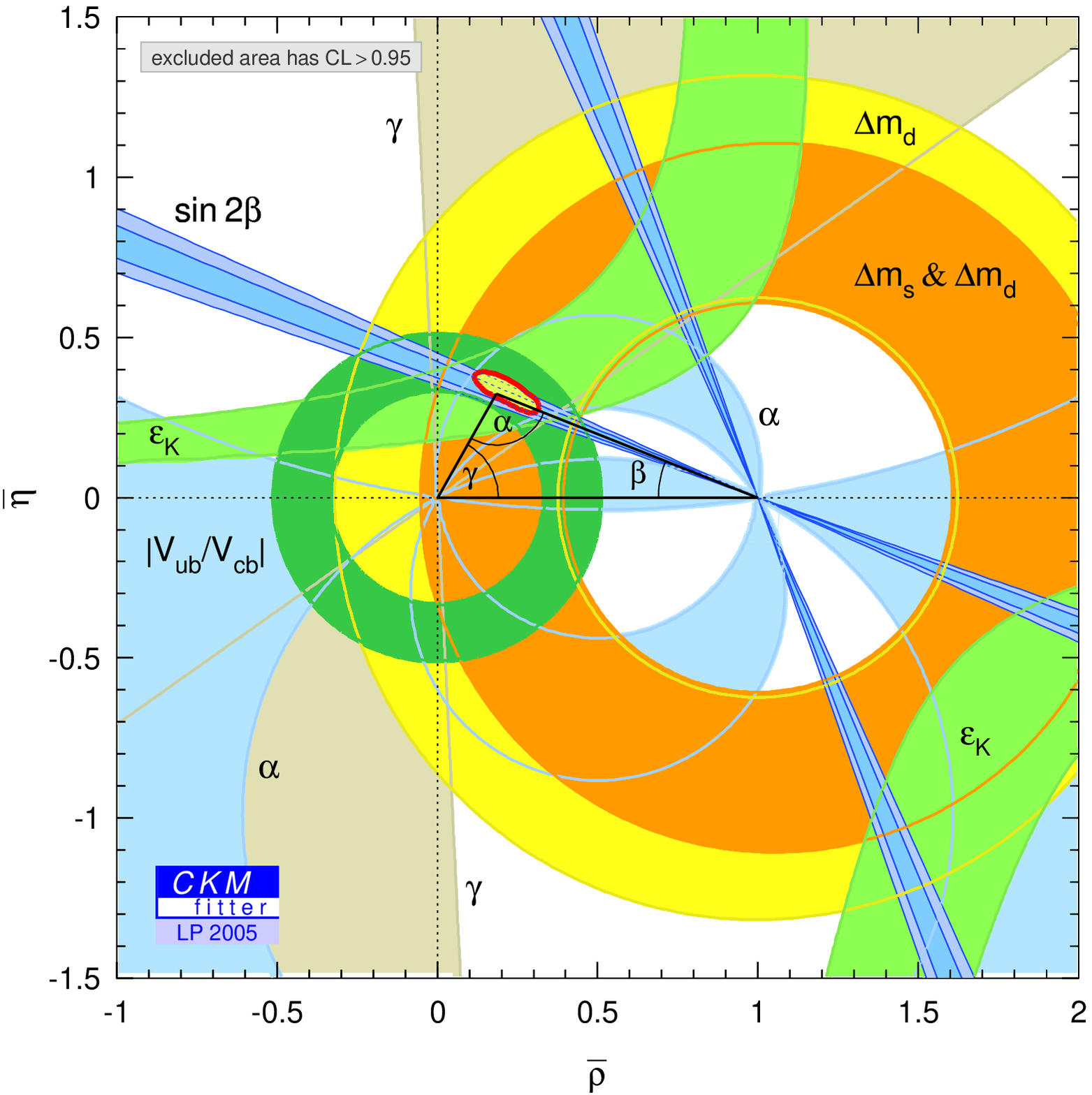}
\epsfxsize0.9\columnwidth
\figurebox{}{}{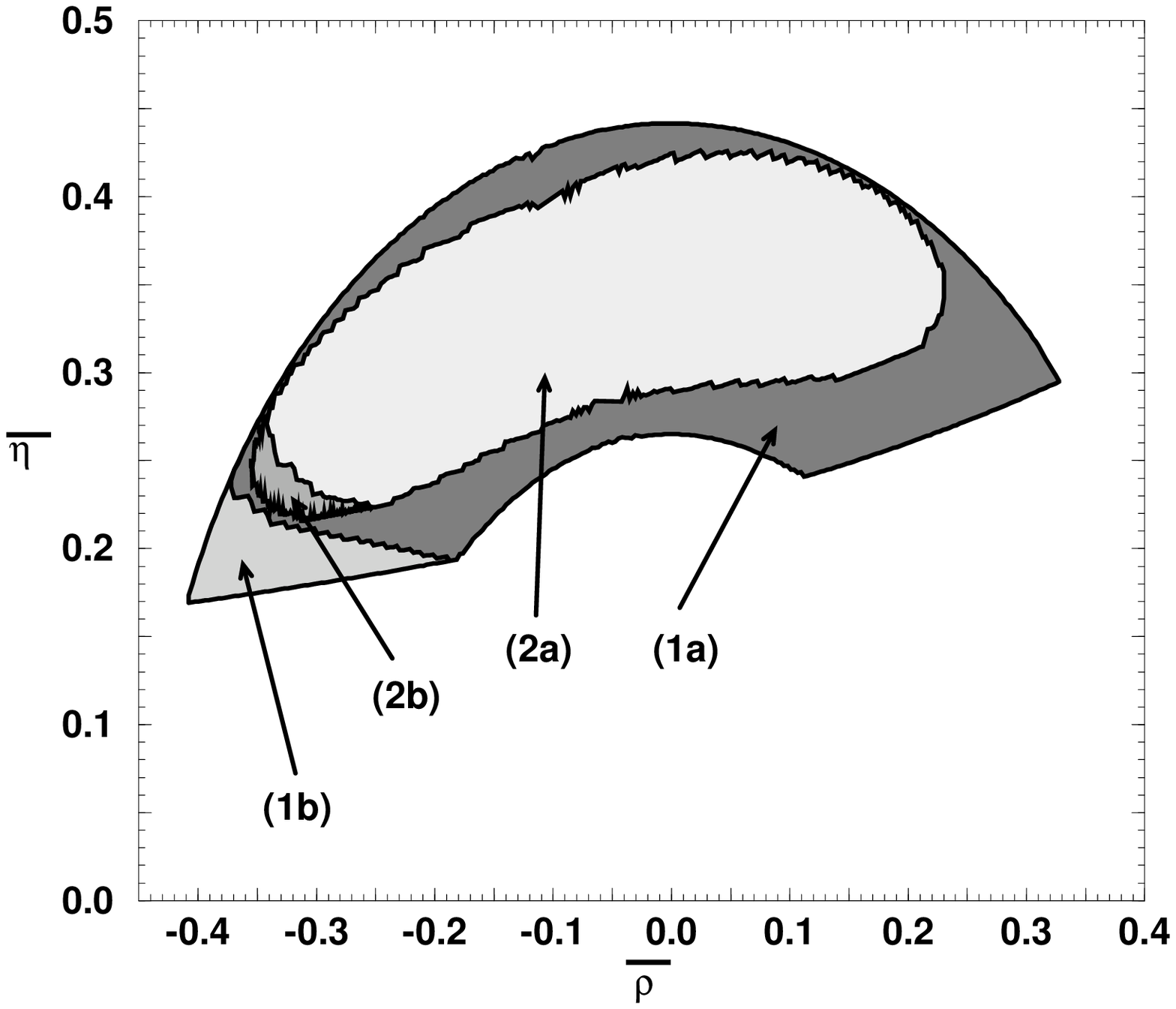}
\caption{LHS: The UT from a global fit to summer 2005 data. RHS: The
  first UT fit using theoretical expressions with NLO QCD corrections, 
  performed in 1995.\protect\cite{hn} At that time only \protect$|V_{ub}|$, 
  \protect$\epsilon_K$
  and \protect$\dm_{B_d}$ could be used. Region 1a corresponds to a scan over 
  1\protect$\sigma$ ranges of the input parameters.}
\label{fig:utres}
\end{figure*}

The result of a global fit of $(\ov{\rho},\ov{\eta})$ to
state-of-the-art summer-2005 data is shown in \fig{fig:utres}.
It uses $\widehat{B}_K = 0.85 \pm 0.02 \pm 0.07$ where the first error
is Gaussian and the second is scanned over according to the standard
CKMfitter method\cite{ckmf}. For the remaining input see\cite{ckmf}.
The fit output is summarised in this table:
{\normalsize
 \begin{tabular}{@{}lcc} \hline &&\\[-0.3cm]
\small  quantity & $\!\!\!\!$\small central $\pm$ ${\rm CL} \equiv 1 \sigma$ &
 \small            $\pm$ ${\rm CL} \equiv 2 \sigma$ \\[0.15cm]
 \hline  && \\[-0.3cm]
 $\bar\rho$                                                             & $    0
.204 ^{\,+    0.035}_{\,-    0.033}$ & $^{\,+    0.095}_{\,-    0.069}$ \\[0.15cm]
 $\bar\eta$                                                             & $    0
.336 ^{\,+    0.021}_{\,-    0.021}$ & $^{\,+    0.045}_{\,-    0.060}$ \\[0.15cm]
 \hline &&\\[-0.3cm]
 $\alpha$~~(deg)                                                        & $
98.4 ^{\,+      6.1}_{\,-      5.6}$ & $^{\,+     16.8}_{\,-     11.8}$ \\[0.15cm]
 $\beta$~~(deg)                                                         & $    2
2.77 ^{\,+     0.87}_{\,-     0.83}$ & $^{\,+     1.92}_{\,-     2.04}$ \\[0.15cm]
$\gamma$~~(deg)                                            & $
58.8 ^{\,+      5.3}_{\,-      5.8}$ & $^{\,+     11.2}_{\,-     15.4}$ \\[0.15cm]
 $|V_{ub}|$~~$[10^{-3}]$                                                & $
3.90 ^{\,+     0.12}_{\,-     0.12}$ & $^{\,+     0.29}_{\,-     0.24}$ 
\\[0.15cm]
 $|V_{td}|$~~$[10^{-3}]$                                                & $
8.38 ^{\,+     0.32}_{\,-     0.44}$ & $^{\,+     0.56}_{\,-     1.29}$ 
\\[0.15cm]
 \hline
 \end{tabular}
The output of the global fit agrees well with the pure tree-level
determinations in Eqs.~(\ref{vub}) and (\ref{gatree}) and \fig{fig:uttree}.

We can use Eq.~(\ref{pdg}) to determine $\theta_{13}= 0.223^\circ\pm
0.007^\circ$ from the fitted $|V_{ub}|$ in the table. With
Eq.~(\ref{pdg}) one finds $\theta_{12}=12.9^\circ\pm 0.1^\circ$ from
\eq{vus} and $\theta_{23}= 2.38^\circ\pm 0.03^\circ $ from \eq{vcb}.
Since $1-\cos \theta_{13}$ is negligibly small, the Wolfenstein
parameters $\lambda$ and $A\lambda^2$ defined in \eq{wex} are simply
given by $V_{us}$ in \eq{vus} and $V_{cb}$ in \eq{vcb}, respectively.

\subsection{$K\to \pi \nu \ov{\nu}$}
The rare decays $\rd K^+ \to \pi^+ \nu \ov{\nu} $ and $\rd K_L \to \pi^0
\nu \ov{\nu} $ provide an {\rd excellent} opportunity to determine the
{\rd unitarity triangle} from $\rd s\to d$ transitions.  With planned
dedicated experiments $\gn (\ov{\rho},\ov{\eta})$ can be determined with
a similar precision as today from $\rd b\to d$ and $\rd b\to u$
transitions at the {\gn B factories}. This is a unique and very powerful
probe of the CKM picture of {\gn FCNCs}.  $\rd Br(K_L \to \pi^0 \nu
\ov{\nu}) $ is proportional to $\gn \ov{\eta}^2$ and dominated by the
top contribution. The theoretical uncertainty of the {\rd
  next-to-leading order (NLO) } prediction\cite{bb1} is {\rd below 2\%}.
$\rd Br(K^+ \to \pi^+ \nu \ov{\nu}) $ defines an ellipse in the $\gn
(\ov{\rho},\ov{\eta})$ plane and has a sizeable charm contribution,
which inflicts a larger theoretical uncertainty on the {\rd
  next-to-leading order (NLO) } prediction\cite{bb2}, leading to $\rd {\cal
  O}(5-10\%)$ uncertainties in extracted CKM parameters.  Parametric
uncertainties from $\rd V_{cb}$ and $\rd m_t$ largely drop out, if $\rd
\sin(2\beta)$ is calculated from $\rd Br(K_L \to \pi^0 \nu \ov{\nu}) $
and $\rd Br(K^+ \to \pi^+ \nu \ov{\nu}) $. Therefore the comparison of
$\sin(2\beta) $ determined in \eq{beta} from the B system with
$\sin(2\beta) $ inferred from $K\to \pi \nu \ov{\nu}$ constitutes a
pristine test of the Standard Model.\cite{bb3} 
The impact of a future 10\% measurements of these rare decay modes on
the UT is shown in \fig{fig:utkp}.

The charm contribution is expanded in two parameters: $\rd m_K^2/m_c^2$ and
$\rd \alpha_s(m_c)$. The calculations of ${\cal O} (m_K^2/m_c^2)$
corrections was recently completed, finding a $7\%$  increase of $\rd
Br(K^+ \to \pi^+ \nu \ov{\nu}) $ with a small residual 
uncertainty.\cite{ims} A new result are the next-to-next-to-leading order
(NNLO) QCD corrections to the charm contribution.\cite{bghn} 
This three-loop calculation
reduces the theoretical error from unknown higher-order terms well below
the parametric uncertainty from $m_c$. The branching ratio
is now predicted as 
\begin{eqnarray}
Br(K^+\to \pi^+ \ov{\nu} \nu) &=& (8.0 \pm 1.1 ) \cdot 10^{-11}. \no
\end{eqnarray}
At {\rd NNLO} one finds the following reduced theoretical uncertainties 
for parameters extracted from $\rd Br(K_L \to \pi^0 \nu \ov{\nu}) $ and
$\rd Br(K^+ \to \pi^+ \nu \ov{\nu}) $:\cite{bghn}
{\rd
\begin{eqnarray}
\frac{\delta |V_{td}|}{|V_{td}|} = 0.010,\;\;
\delta \sin(2\beta) = 0.006, && 
\delta \gamma = 1.2^\circ \no
\end{eqnarray}
\begin{figure}
\epsfxsize\columnwidth
\figurebox{}{}{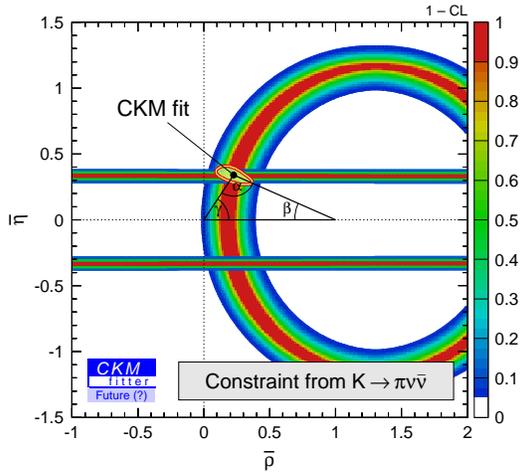}
\caption{Unitarity triangle from a future measurement of  
  \protect$\rd Br(K_L \to \pi^0 \nu \ov{\nu}) $ and \protect$\rd Br(K^+
  \to \pi^+ \nu \ov{\nu}) $ with 10\% accuracy. The current UT from the
  2005 global fit \protect\cite{ckmf} is overlaid. The comparison of
  this UT constructed from $s\to d$ FCNCs with the UT found from b
  decays probes the CKM origin of FCNCs precisely and in a unique way.}
\label{fig:utkp}
\end{figure}

\section{CP violation in $b\to s$ penguin decays}
Within the Standard Model the mixing-induced CP asymmetries in $\rd b\to
s \ov{q} q$ penguin amplitudes are proportional to $\sin(2\beta)^{\rm
  eff}$ which equals $\sin(2 \beta)$ in \eq{beta} up to small
corrections from a penguin loop with an up quark. In $\rd b\to s \ov{u}
u$ decays there is also a color-suppressed tree amplitude. In any case
the corrections are parametrically suppressed by $ |
V_{ub}V_{us}/(V_{cb}V_{cs} ) | \sim 0.025$. The experimental situation
is shown in \fig{fig:cpp}.
\begin{figure}
\epsfxsize1\columnwidth
\figurebox{}{}{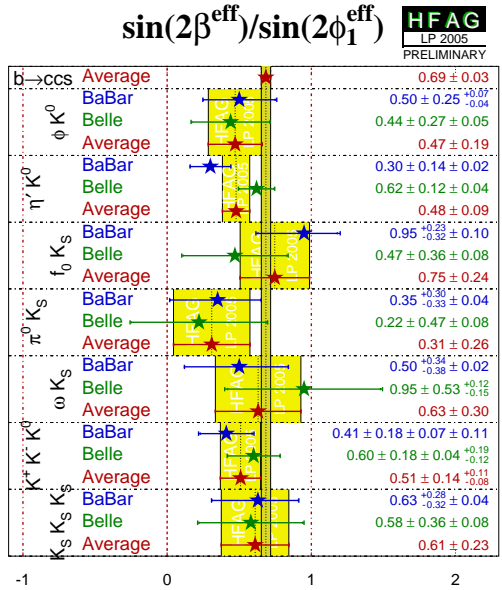}
\caption{\protect $\sin(2 \beta)^{\rm
  eff}$ from various penguin decays. The small vertical yellow band is 
  \protect $\sin(2 \beta) $ from \protect\eq{beta}.\protect\cite{hfag}}
\label{fig:cpp}
\end{figure}
A naive average of the measurements of \fig{fig:cpp} gives 
\begin{eqnarray}
\sin(2\beta)^{\rm eff} =0.51\pm 0.06, \no
\end{eqnarray}
which is below the value of $\sin(2 \beta)$ from tree-level $\rd b\to \ov{c} c
s$ decays in \eq{beta} by 3$\sigma$. Moreover QCD factorisation finds a
small and positive correction to $\rd \sin (2\beta_{\rm eff})\!-\!\sin
(2\beta)$ from up-quark effects.\cite{b} While the significance of the
deviation has decreased since the winter 2005 conferences, the
mixing-induced CP asymmetries in $\rd b\to s \ov{q} q$ decays stay interesting
as they permit large effects from new physics. While in $B_d$ decays the
needed interference of a $B_d$ and $\Bbar_d$ decay to the same final
state requires a neutral K meson in the final state, $b\to s \ov{q} q$
decays of $B_s$ mesons go to a flavourless $ \ov{s} s \ov{q} q$
state, so that the desired CP effects can be studied in any final state.
Hence $B_s$ physics has the potential to become the ``El Dorado'' of
$b\to s\ov{q} q$ penguin physics.

\section*{Acknowledgments}
This presentation was made possible through the input and help from
Martin Beneke, Ed Blucher, Tom Browder, Oliver Buchm\"uller, Christine
Davies, Henning Ulrik Fl\"acher, Paolo Franzini, Tim Gershon, Martin
Gorbahn, Ulrich Haisch, Christopher Hearty, Matthias Jamin, Bob
Kowalewski, Heiko Lacker, Zoltan Ligeti, Antonio Limosani, Mike Luke,
Matthias Neubert, Maurizio Pierini, Jim Smith, Iain Stewart, St\'ephane
T'Jampens, Nikolai Uraltsev and Matthew Wingate. I am especially
grateful to Vincenzo Cirigliano and Bj\"orn Lange for their thorough and
patient explanations of recent progress in the fields of chiral
perturbation theory and B meson shape functions. Very special thanks go
to Andreas H\"ocker and Heiko Lacker for their day-and-night work on the
presented CKMfitter plots.

\section*{DISCUSSION}

\begin{description}
\item[Luca Silvestrini] (Rome and Munich):\\
  Maybe rather than saying that QCD cannot explain $\sin{2\beta}^{\rm
    eff}$ in $b \to s $ penguins, one should say that a particular model
  of power supressed corrections due to Beneke and Co.\ cannot do it,
  but this is not a model-independent statement. If you just want to use
  data, and you say you do not know anything about power corrections, I
  do not think that you can infere anything from that plot.
  
\item[Ulrich Nierste{\rm :}] The parametric suppression of the up-quark
  pollution by $ |V_{ub}V_{us}/(V_{cb}V_{cs} ) | \sim 0.025$ is
  undisputed. Further the leading term in the $1/m_b$ expansion of
  $\sin{2\beta}^{\rm eff} -\sin(2\beta) $ can be reliably computed and
  results in the finding of Ref.\cite{b} that $\sin{2\beta}^{\rm eff}
  -\sin(2\beta) $ is small and positive for the measured modes. It is
  true that the size of the modeled power corrections is currently
  widely debated. Yet I am not aware of any possible dynamical QCD
  effect in two-body B decays which is formally ${\cal O}(1/m_b)$, 
  large in magnitude and further comes with the large strong phase 
  needed to flip the sign of $\sin{2\beta}^{\rm eff} -\sin(2\beta) $.

\end{description}


{\small
\begin{thebibliography}{99}
  
\bibitem{pdg} S.~Eidelman {\it et al.}  [Particle Data Group],
  {\it Phys.\ Lett.} B {\bf 592}, 1 (2004).

\bibitem{w}   L.~Wolfenstein,
  Phys.\ Rev.\ Lett.\  {\bf 51}, 1945 (1983).

\bibitem{blo} A.~J.~Buras,
  M.~E.~Lautenbacher, G.~Ostermaier, \prd\ {\bf 50}, 3433 (1994).

\bibitem{hfag} \emph{Heavy Flavor Averaging Group}, 
  {\tt\footnotesize http://www.slac.stanford.edu/xorg/hfag}
\bibitem{ckmf}   A.~H\"ocker, H.~Lacker, S.~Laplace and F.~Le Diberder,
  Eur.\ Phys.\ J.\ C {\bf 21}, 225 (2001)
  [arXiv:hep-ph/0104062].
   J.~Charles {\it et al.}  [CKMfitter Group],
  Eur.\ Phys.\ J.\ C {\bf 41}, 1 (2005)
  [arXiv:hep-ph/0406184].
  {\tt\footnotesize http://www.slac.stanford.edu/xorg/ckmfitter}
\bibitem{utf}
   \mbox{M.\ Ciuchini} {\it et al.},
  JHEP {\bf 0107}, 013 (2001)[arXiv:hep-ph/0012308].
  M.~Bona {\it et al.}  [UTfit Collaboration],
  arXiv:hep-ph/0509219.
  {\tt\footnotesize http://utfit.roma1.infn.it}


\bibitem{s} Iain Stewart, these proceeedings.  

\bibitem{h} 
  John Hardy, talk at {\it KAON 2005 Int.\ Workshop}, Jul 13-17, 2005, 
  Evanston, USA.

\bibitem{ag}   M.~Ademollo and R.~Gatto,
  {\it Phys.\ Rev.\ Lett.} {\bf 13}, 264 (1964).

\bibitem{cms}
 A.~Czarnecki, W.~J.~Marciano and A.~Sirlin,
  Phys.\ Rev.\ D {\bf 70}, 093006 (2004)
  [arXiv:hep-ph/0406324].

\bibitem{c} 
  Vincenzo Cirigliano, 
  talk at {\it KAON 2005 Int.\ Workshop}, Jul 13-17, 2005, 
  Evanston, USA, and private communication.

\bibitem{gl}
  J.~Gasser and H.~Leutwyler,
  Annals Phys.\  {\bf 158}, 142 (1984).

\bibitem{em}
 V.~Cirigliano, M.~Knecht, H.~Neufeld, H.~Rupertsberger and P.~Talavera,
  Eur.\ Phys.\ J.\ C {\bf 23}, 121 (2002)
  [arXiv:hep-ph/0110153].
  V.~Cirigliano, H.~Neufeld and H.~Pichl,
  Eur.\ Phys.\ J.\ C {\bf 35}, 53 (2004)
  [arXiv:hep-ph/0401173].
   T.~C.~Andre,
  arXiv:hep-ph/0406006.
  S.~Descotes-Genon and B.~Moussallam,
  arXiv:hep-ph/0505077.

\bibitem{ff}  
   H.~Leutwyler and M.~Roos,
  Z.\ Phys.\ C {\bf 25}, 91 (1984).
    P.~Post and K.~Schilcher,
  Nucl.\ \mbox{Phys.\ B {\bf 599}, 30 (2001)}
  [arXiv:hep-ph/0007095].\nolinebreak
 J.~Bijnens and P.~Talavera,
  \mbox{Nucl.~Phys.~B {\bf 669}, 341 (2003)}
  [arXiv:hep-ph/0303103].
      M.~Jamin, J.~A.~Oller and A.~Pich,
  JHEP {\bf 0402}, 047 (2004)
  [arXiv:hep-ph/0401080].
  V.~Cirigliano,
  Int.\ J.\ Mod.\ Phys.\ A {\bf 20}, 3732 (2005).\\
  $f_+^{K^0\pi^-}$ was computed in quenched lattice QCD in:
  D.~Becirevic {\it et al.},
  Eur.\ Phys.\ J.\ A {\bf 24S1}, 69 (2005)
  [arXiv:hep-lat/0411016];
  V.~Lubicz {\it et al.},
  presented at {\it DAFNE 2004: Workshop on Physics at Meson Factories, 
   Rome, Frascati, Italy, 7-11 Jun 2004}


\bibitem{m} 
  W.~J.~Marciano,
  {\it Phys.\ Rev.\ Lett.}  {\bf 93}, 231803 (2004)
  [arXiv:hep-ph/0402299].

\bibitem{milc} 
  C.~Aubin {\it et al.}  [MILC Collaboration],
  {\it Nucl.\ Phys.\ Proc.\ Suppl.}  {\bf 140}, 231 (2005)
  [arXiv:hep-lat/0409041].

\bibitem{pp}
  A.~Pich and J.~Prades,
  JHEP {\bf 9806}, 013 (1998)
  [arXiv:hep-ph/9804462].
  E.~Gamiz, M.~Jamin, A.~Pich, J.~Prades and F.~Schwab,
  JHEP {\bf 0301}, 060 (2003)
  [arXiv:hep-ph/0212230].

\bibitem{gjpps}
  E.~Gamiz, M.~Jamin, A.~Pich, J.~Prades and F.~Schwab,
  Phys.\ Rev.\ Lett.\  {\bf 94}, 011803 (2005)
  [arXiv:hep-ph/0408044].

\bibitem{opal}
 G.~Abbiendi {\it et al.}  [OPAL Collaboration],
  Eur.\ Phys.\ J.\ C {\bf 35}, 437 (2004)
  [arXiv:hep-ex/0406007].

\bibitem{sch}
  K.~R.~Schubert,
  Int.\ J.\ Mod.\ Phys.\ A {\bf 19}, 1004 (2004).

\bibitem{sv}
  M.~A.~Shifman and M.~B.~Voloshin,
  Sov.\ J.\ Nucl.\ Phys.\  {\bf 41}, 120 (1985)
  [Yad.\ Fiz.\  {\bf 41}, 187 (1985)].
  I.~I.~Y.~Bigi, N.~G.~Uraltsev and A.~I.~Vainshtein,
  Phys.\ Lett.\ B {\bf 293}, 430 (1992)
  [Erratum-ibid.\ B {\bf 297}, 477 (1993)]
  [arXiv:hep-ph/9207214].

\bibitem{gu}
  P.~Gambino and N.~Uraltsev,
  Eur.\ Phys.\ J.\ C {\bf 34}, 181 (2004)
  [arXiv:hep-ph/0401063].
  C.~W.~Bauer, Z.~Ligeti, M.~Luke, A.~V.~Manohar and M.~Trott,
  Phys.\ Rev.\ D {\bf 70}, 094017 (2004)
  [arXiv:hep-ph/0408002].

\bibitem{bf}
  O.~Buchm\"uller and H.~Fl\"acher,
  arXiv:hep-ph/0507253, and references therein.

\bibitem{bm} 
  C.~W.~Bauer and A.~V.~Manohar,
  Phys.\ Rev.\ D {\bf 70}, 034024 (2004)
  [arXiv:hep-ph/0312109].
  S.~W.~Bosch, B.~O.~Lange, M.~Neubert and G.~Paz,
  Phys.\ Rev.\ Lett.\  {\bf 93}, 221801 (2004)
  [arXiv:hep-ph/0403223].

\bibitem{ls} 
    K.~S.~M.~Lee and I.~W.~Stewart,
  Nucl.\ Phys.\ B {\bf 721}, 325 (2005)
  [arXiv:hep-ph/0409045].
   S.~W.~Bosch, M.~Neubert and G.~Paz,
  JHEP {\bf 0411}, 073 (2004)
  [arXiv:hep-ph/0409115].
  M.~Beneke, F.~Campanario, T.~Mannel and B.~D.~Pecjak,
  JHEP {\bf 0506}, 071 (2005)
  [arXiv:hep-ph/0411395].


\bibitem{lnp}
  B.~O.~Lange, M.~Neubert and G.~Paz,
  arXiv:hep-ph/0504071.

\bibitem{mr}  
    T.~Mannel and S.~Recksiegel,
  Phys.\ Rev.\ D {\bf 63}, 094011 (2001)
  [arXiv:hep-ph/0009268].


\bibitem{lr}
  A.~K.~Leibovich, I.~Low and I.~Z.~Rothstein,
  Phys.\ Lett.\ B {\bf 513}, 83 (2001)
  [arXiv:hep-ph/0105066].
  A.~K.~Leibovich, I.~Low and I.~Z.~Rothstein,
  Phys.\ Rev.\ D {\bf 61}, 053006 (2000)
  [arXiv:hep-ph/9909404].
  A.~K.~Leibovich, I.~Low and I.~Z.~Rothstein,
  Phys.\ Lett.\ B {\bf 486}, 86 (2000)
  [arXiv:hep-ph/0005124].

\bibitem{cvub}
  A.~Bornheim {\it et al.}  [CLEO Collaboration],
  Phys.\ Rev.\ Lett.\ {\bf 88}, 231803 (2002) [arXiv:hep-ex/0202019].
  
\bibitem{bevub}
  H.~Kakuno {\it et al.}  [BELLE Collaboration],
  Phys.\ Rev.\ Lett.\  {\bf 92}, 101801 (2004)
  [arXiv:hep-ex/0311048].
  I.~Bizjak {\it et al.}  [Belle Collaboration],
  arXiv:hep-ex/0505088.
  A.~Limosani {\it et al.}  [Belle Collaboration],
  Phys.\ Lett.\ B {\bf 621}, 28 (2005)
  [arXiv:hep-ex/0504046].
\bibitem{bavub}
  B.~Aubert {\it et al.}  [BaBar Collaboration],
  arXiv:hep-ex/0408075.
  B.~Aubert {\it et al.}  [BABAR Collaboration],
  Phys.\ Rev.\ Lett.\  {\bf 95}, 111801 (2005)
  [arXiv:hep-ex/0506036].
  B.~Aubert {\it et al.}  [BABAR Collaboration],
  arXiv:hep-ex/0509040. (The average quoted in \eq{vub} uses a slightly
  different, preliminary result.)

\bibitem{glw}  
  M.~Gronau and D.~London.,
  Phys.\ Lett.\ B {\bf 253}, 483 (1991).
  M.~Gronau and D.~Wyler,
  Phys.\ Lett.\ B {\bf 265}, 172 (1991).

\bibitem{ggssz}
  M.~Gronau, Y.~Grossman, N.~Shuhmaher, A.~Soffer and J.~Zupan,
  Phys.\ Rev.\ D {\bf 69}, 113003 (2004)
  [arXiv:hep-ph/0402055].

\bibitem{glo}
  M.~Gronau and D.~London,
  Phys.\ Rev.\ Lett.\  {\bf 65}, 3381 (1990).

\bibitem{hpqcdbk}
  E.~Gamiz, S.~Collins, C.~T.~H.~Davies, J.~Shigemitsu and M.~Wingate,
  arXiv:hep-lat/0509188.

\bibitem{bgm}
  T.~Becher, E.~Gamiz and K.~Melnikov,
  arXiv:hep-lat/0507033.

\bibitem{twb} 
  B.~Aubert {\it et al.}  [BABAR Collaboration],
  Phys.\ Rev.\ Lett.\  {\bf 94}, 161803 (2005)
  [arXiv:hep-ex/0408127].
  K.~Abe {\it et al.}  [Belle Collaboration],
  arXiv:hep-ex/0507037.
  R.~Barate {\it et al.}  [ALEPH Collaboration],
  Phys.\ Lett.\ B {\bf 492}, 259 (2000)
  [arXiv:hep-ex/0009058].
  K.~Ackerstaff {\it et al.}  [OPAL collaboration],
  Eur.\ Phys.\ J.\ C {\bf 5}, 379 (1998)
  [arXiv:hep-ex/9801022].
  T.~Affolder {\it et al.}  [CDF Collaboration],
  Phys.\ Rev.\ D {\bf 61}, 072005 (2000)
  [arXiv:hep-ex/9909003].

\bibitem{hpqcdfb}
  A.~Gray {\it et al.}  [HPQCD Collaboration],
  arXiv:hep-lat/0507015.

\bibitem{hn}
  S.~Herrlich and U.~Nierste,
  Phys.\ Rev.\ D {\bf 52}, 6505 (1995)
  [arXiv:hep-ph/9507262].

\bibitem{bb1}
  G.~Buchalla and A.~J.~Buras,
  Nucl.\ Phys.\ B {\bf 400}, 225 (1993).
  G.~Buchalla and A.~J.~Buras,
  M.~Misiak and J.~Urban,
  Phys.\ Lett.\ B {\bf 451}, 161 (1999)
  [arXiv:hep-ph/9901278].
  Nucl.\ Phys.\ B {\bf 548}, 309 (1999) [arXiv:hep-ph/9901288].

\bibitem{bb2}
  G.~Buchalla and A.~J.~Buras,
  Nucl.\ Phys.\ B {\bf 412}, 106 (1994)
  [arXiv:hep-ph/9308272].

\bibitem{bb3}
  G.~Buchalla and A.~J.~Buras,
  Phys.\ Lett.\ B {\bf 333}, 221 (1994)
  [arXiv:hep-ph/9405259].

\bibitem{ims}  
  G.~Isidori, F.~Mescia and C.~Smith,
  Nucl.\ Phys.\ B {\bf 718}, 319 (2005)
  [arXiv:hep-ph/0503107].
  A.~F.~Falk, A.~Lewandowski and A.~A.~Petrov,
  Phys.\ Lett.\ B {\bf 505}, 107 (2001)
  [arXiv:hep-ph/0012099].

\bibitem{bghn}
  A.~J.~Buras, M.~Gorbahn, U.~Haisch and U.~Nierste,
  arXiv:hep-ph/0508165.

\bibitem{b}  
  M.~Beneke,
  Phys.\ Lett.\ B {\bf 620}, 143 (2005)
  [arXiv:hep-ph/0505075].



\end{thebibliography}
}
\end{document}